\documentclass[aps,pre,amsmath,amssymb,floatfix, showkeys, showpacs]{revtex4-2}

\usepackage{graphicx}
\usepackage{caption}
\usepackage{subcaption}
\usepackage{amsmath}
\usepackage{amssymb}
\usepackage{xcolor}

\begin{document}

\title{$\mathcal{PT}$-symmetric branched optical lattices:\\ 
Spectral properties and stability of solitons}

\author{O.K. Tojakhmadova, 
T. Akhmadjanov, 
M.E. Akramov}

\affiliation{National University of Uzbekistan, Universitet Str. 4, Tashkent 100174, Uzbekistan}

\begin{abstract}
We propose a model for tunable $\mathcal{PT}$-symmetric branched optical lattices by investigating both linear and nonlinear Schrödinger equations with a $\mathcal{PT}$-symmetric periodic potential on the graph and solving them by imposing weighted vertex boundary conditions.
A constraint derived from these vertex conditions determines the exceptional point of the system. 
In the $\mathcal{PT}$ unbroken phase, this constraint enforces $\mathcal{PT}$-symmetric boundary conditions at the vertices, ensuring a purely real spectrum; its violation leads to the emergence of complex eigenvalues in the linear regime. In the nonlinear regime, the same constraint determines the linear stability of solitons: satisfying the constraint yields stable solitons, whereas violating it corresponds to unstable solitons. 
\end{abstract}



\maketitle

\section{Introduction}
Non-Hermitian $\mathcal{PT}$-symmetric Hamiltonians have attracted much attention over the past decades since the pioneering paper by Bender \textit{et al}~\cite{Bender0}. As demonstrated in that work, it is well established that a $\mathcal{PT}$-symmetric Hamiltonian can possess real eigenvalues when the $\mathcal{PT}$-symmetry is unbroken, even though it is non-Hermitian. One of the most effective ways to construct such a $\mathcal{PT}$-symmetric Hamiltonian is to add a $\mathcal{PT}$-symmetric complex potential whose real part is an even function and imaginary part is an odd function.

Beyond quantum mechanics,
$\mathcal{PT}$-symmetric systems have been widely studied in various physical contexts, from topological phases to biological systems.
Among these, one of the most widely studied fields is optical/photonic systems (see, for example,~\cite{pt_opt1, pt_opt2, pt_opt3}).
In linear optical media, $\mathcal{PT}$-symmetry appears in structures with a spatially symmetric refractive index and balanced gain–loss regions.
When $\mathcal{PT}$-symmetry is broken, the intensity of the beam grows or decreases exponentially. This feature allows us to create light-controlled devices~\cite{light_control}.
In these studies~\cite{band1, Musslimani1, Musslimani2}, 
the band structures of $\mathcal{PT}$-symmetric optical lattices have been investigated, 
revealing how different gain–loss modulations give rise to distinct spectral features and 
$\mathcal{PT}$-symmetry-breaking behaviors.
The first experimental realization of $\mathcal{PT}$-symmetric optical systems was achieved by constructing passive \cite{zero_PT} and active \cite{first_PT} configurations exhibiting $\mathcal{PT}$-symmetry breaking between two channels.
Unidirectional invisible media was created using synthetic photonic lattices \cite{Peschel}.
Two coupled microcavity resonators have been shown to demonstrate $\mathcal{PT}$-symmetry breaking, and nonlinear effects provide nonreciprocity in next-generation $\mathcal{PT}$-symmetric optical systems \cite{Peng}.
Optical isolation \cite{Chang}, phase transitions \cite{atomic} and topological edge states \cite{edge_state} are demonstrated experimentally in microresonators, atomic lattices and quantum walks, respectively.

Theoretically, its application to nonlinear wave phenomena is interesting in the context of the rich interplay between $\mathcal{PT}$-symmetry and nonlinearity~\cite{Musslimani1, Musslimani2, Nixon, Musslimani, Yang1,Barashenkov,  Yan, Bakirtas1,  Malomed1, Nixon2,  Nixon3, Yang2, Yang3,  Holger, Midya, Nixon4, Mihalache, Bakirtas2}.
Especially, various properties of $\mathcal{PT}$-symmetric solitons in the nonlinear Schrödinger case have been studied in the literature.
The linear stability of $\mathcal{PT}$-symmetric solitons with periodic potentials in both one- and two-dimensional systems has been studied in $\mathcal{PT}$ optical lattices~\cite{Nixon,Musslimani,Yang1}.
The anomalous behaviour occurring for larger gain–loss amplitudes was investigated in~\cite{Barashenkov}.
Exact analytical solutions for the nonlinear Schrödinger and Burgers equations in one and two dimensions were obtained for different types of $\mathcal{PT}$-symmetric potentials~\cite{Yan}.
An exact analytical solution for the Scarf-II potential with cubic–quintic nonlinearity terms was analyzed ~\cite{Bakirtas1}.
An extension of that work to metric graphs for cubic nonlinearity, together with linear stability, was discussed in~\cite{Old}.
Existence of optical solitons in $\mathcal{PT}$-symmetric lattices was presented experimentally in synthetic dimensions \cite{Solitons}.

A new integrable nonlocal nonlinear Schrödinger equation with a self-induced $\mathcal{PT}$-symmetric potential and its discrete model were introduced in~\cite{Ablowitz1, Ablowitz2}.
Their extension to metric graphs was discussed, respectively, in~\cite{Mashrab1, Mashrab2}.
Nonlinear evolution equations are considered in branched structures, where the branched structure is modeled in terms of metric graphs~\cite{Zarif, zar2011, Adami, noja, Our2015, DP2015, Adami16, dimarecent, SGN2020}.
Transparent boundary conditions for the nonlocal nonlinear Schrödinger equation have been developed both on the line~\cite{TBC_nnlse1} and on graphs~\cite{TBC_nnlse2}.

In this paper, we consider the $\mathcal{PT}$-symmetric branched optical lattice in both linear and nonlinear regimes.
The $\mathcal{PT}$-symmetric branched optical lattice is modeled using a $\mathcal{PT}$-symmetric periodic potential on a metric graph.
In the linear regime, we obtain the dispersion relation using the plane-wave expansion method by imposing boundary conditions with some weight at the branching point, called the vertex.
Violation of the constraint on the weight parameters, due to improper choice of their values, leads to the breaking of $\mathcal{PT}$-symmetry, resulting in complex eigenvalues.
This violation appears in the dispersion relation as a bifurcation of eigenvalues.


In the nonlinear regime, we investigate the nonlinear Schrödinger equation with the $\mathcal{PT}$-symmetric periodic potential on the graph.
The linear stability of $\mathcal{PT}$-symmetric solitons on the graph is explored using the perturbation of the complex field amplitude.

The proposed model exhibits a tunable $\mathcal{PT}$-symmetric phase transition, where the stability of the $\mathcal{PT}$-symmetric system is governed by the coefficients of the weighted boundary conditions.
Near the exceptional point, modulating the these coefficients induces $\mathcal{PT}$-symmetry breaking in the system, allowing for dynamic control over the system's phase.

Experimentally, such a structure can be realized via femtosecond laser micromachining, where high-intensity laser pulses are focused into a dielectric crystal to induce localized changes in the refractive index, thereby forming an integrated network of intersecting waveguides. The $\mathcal{PT}$-symmetric potential is established in such a way that the complex refractive index profile ensures a spatially balanced distribution of gain and loss by satisfying the condition $n(x)=n^*(-x)$.


This paper is organized as follows. In the next section, we briefly mention and demonstrate some new results of the linear Schrödinger equation with a $\mathcal{PT}$ periodic potential on a line.
Section~\ref{Branched_PT_lattices} is devoted to obtaining the dispersion relation of the branched $\mathcal{PT}$ optical lattice.
$\mathcal{PT}$ solitons are considered by adding a cubic nonlinear term to the Schrödinger equation on a line and on the graph, respectively, in Sections~\ref{NLSE_on_line} and~\ref{NLSE_on_star}.
Finally, Section~\ref{conclusion} provides some concluding remarks.

\section{$\mathcal{PT}$-symmetric lattices}

In this section, we consider $\mathcal{PT}$-symmetric optical lattice by following the paper by Nixon \textit{et al}~\cite{Nixon}.
The linear Schrödinger equation with a $\mathcal{PT}$-symmetric periodic potential describing a 1D $\mathcal{PT}$ lattice is given by
\begin{eqnarray}\label{LSE1}
    i\partial_t \Psi(x,t) + \partial_x^2 \Psi(x,t) + V(x)\Psi(x,t) = 0,
\end{eqnarray}
where the $\mathcal{PT}$-symmetric periodic potential is defined as
\begin{eqnarray}\label{potential}
    V(x) = V_0(\cos^2(x)+i W_0 \sin(2x)),
\end{eqnarray}
with $V_0$ and $W_0$ being real constants.
This potential represents a prototypical $\mathcal{PT}$-symmetric periodic structure widely used in optics. 
The real part corresponds to a periodic modulation of the refractive index, which can be realized 
experimentally by interfering laser beams. The imaginary part 
describes a balanced gain/loss profile, achievable in waveguides with alternating 
amplification and absorption regions. Such systems have important applications in optoelectronic/photonic devices, including optical switching, mode selection, and enhanced sensing near exceptional points.

Using the separation of variables
\begin{eqnarray*}
    \Psi(x,t)=e^{-i\mu t}\psi(x)
\end{eqnarray*}
one obtains the stationary version of Eq.~\eqref{LSE1} in the form
\begin{eqnarray}\label{LSE2}
    \mu \psi(x)+\partial_x^2 \psi(x)+V(x)\psi(x)=0.
\end{eqnarray}

Since in the following section, we consider branched $\mathcal{PT}$-symmetric lattices, where the boundary conditions must be introduced effectively, it is convenient to use the plane-wave expansion approach. This method provides a flexible framework for constructing the dispersion relation and can be readily adapted to systems with complex lattice connectivity.

The $\mathcal{PT}$-symmetric potential can be rewritten using Euler’s formula as
\begin{equation}\label{EP}
    V(x)=\frac{V_0}{2}+\frac{V_0}{4}\left(1+2W_0\right)e^{2ix}+\frac{V_0}{4}\left(1-2W_0\right)e^{-2ix}.
\end{equation}
Formally, this plane wave expansion method leads to an infinite-dimensional Hamiltonian, but the essential features of the wave function can be captured by a finite number of Fourier modes. Thus, we truncate the system to $n\in[-N,N]$, yielding a finite-dimensional Hamiltonian matrix.

The plane-wave expansion of the wave function is written as
\begin{eqnarray}\label{Psi2}
    \psi(x)=\sum_{n=-N}^{+N} {\varphi_n e^{i(k+G_n)x}},
\end{eqnarray}
where $\varphi_n$ is the amplitude of each mode and $G_n=2n$ due to the periodicity of the potential, which is $\pi$.

Projecting onto $\varphi_n$, we obtain
\begin{eqnarray}\label{eig1}
    \left[(k+G_n)^2 -\frac{V_0}{2} \right]\varphi_n-\frac{V_0}{4}\left(1+2W_0\right) \varphi_{n-1}
    -\frac{V_0}{4}\left(1-2W_0\right)\varphi_{n+1}
    =\mu \varphi_n.
\end{eqnarray}
Eq.~\eqref{eig1} can be written as the matrix eigenvalue problem
\begin{eqnarray}
    H(k)\varphi=\mu \varphi,
\end{eqnarray}
where $\varphi=(\varphi_{-N},...,\varphi_N)^T$ and $(2N+1)\times (2N+1)$ tridiagonal Hamiltonian $H(k)$ is defined as
\begin{align}\label{H1}
H_{n,m}(k)=\left\{ 
\begin{matrix}
    (k+G_n)^2 -\frac{V_0}{2}, & n=m, \\[1ex]
    -\frac{V_0}{4}\left(1-2W_0\right), & n=m- 1, \\[1ex]
    -\frac{V_0}{4}\left(1+2W_0\right), & n=m+1, \\[1ex]
    0, & |n-m|>1,
\end{matrix} \right.,
\end{align}
where $-N\leq n,m \leq N$.

Here, we use a similarity transformation to determine the condition ensuring that the eigenvalue $\mu$ remains real. Under this condition, the Hamiltonian $H(k)$ can be mapped to its Hermitian counterpart.

Let $S$ be a diagonal matrix of the same size as $H(k)$ with entries $S_{n,n}=r^n$, where $r$ is a constant.
We apply the similarity transformation to $H(k)$ given by
\begin{eqnarray}
    \mathcal{H}(k) = S^{-1}H(k)S.
\end{eqnarray}
The off-diagonal elements satisfy
\begin{align*}
    \mathcal{H}_{n,n+1}(k) & =S^{-1}_{n,n}H_{n,n+1}(k) S_{n+1,n+1} = rH_{n,n+1}(k), \\
    \mathcal{H}_{n+1,n}(k) & =S^{-1}_{n+1,n+1}H_{n+1,n}(k) S_{n,n} = r^{-1}H_{n+1,n}(k).
\end{align*}
The eigenvalue $\mu$ is real if and only if the transformed Hamiltonian $\mathcal{H}(k)$ is Hermitian, which requires $\mathcal{H}_{n,n+1}(k)=\mathcal{H}^*_{n+1,n}(k)$, yielding
\begin{eqnarray}
    r=
    e^{i\theta} \sqrt{\frac{1+2W_0}{1-2W_0}},
\end{eqnarray}
where $\theta$ is an arbitrary real constant.

The resulting Hamiltonian $\mathcal{H}$ is a tridiagonal matrix with entries
\begin{align}\label{Hnew}
\mathcal{H}_{n,m}(k)=\begin{cases} 
    (k+G_n)^2 -\dfrac{V_0}{2}, & n=m, \\[1ex]
    -\dfrac{V_0}{4} e^{\mp i\theta} \sqrt{1-4W_0^2}, & n=m\pm 1, \\[1ex]
    0, & |n-m|>1.
\end{cases}
\end{align}

The Hermiticity of the Hamiltonian $\mathcal{H}(k)$ is broken if the expression under the square root in the off-diagonal elements becomes negative. Therefore, the condition for the eigenvalues to remain real is  
\begin{align}\label{cond}
    |W_0| \leq 0.5.
\end{align}

Note that the boundary values of $W_0$ in Eq.~\eqref{cond} define the phase transition points between real and complex eigenvalues.

\begin{figure}[t!]
    \centering
    \begin{minipage}{0.4\textwidth}
        \centering
    \includegraphics[width=0.99\linewidth]{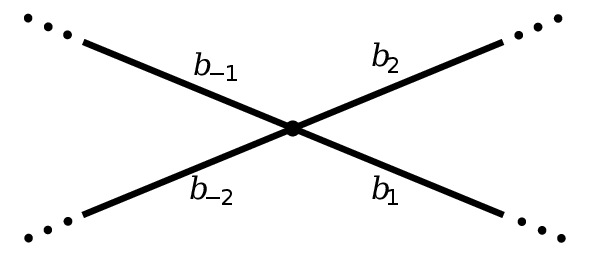}
     \vspace{4mm}
    \small (a)
    \end{minipage}
    \begin{minipage}{0.4\textwidth}
        \centering
        \includegraphics[width=0.99\linewidth]{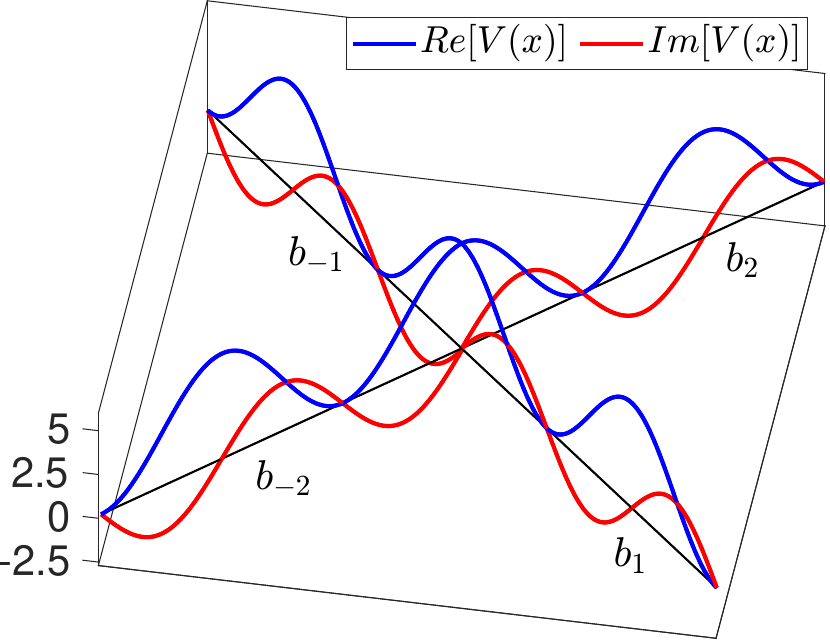}
         \vspace{2mm}
    \small (b)
    \end{minipage}
\caption{Basic star graph with four edges (panel a), and real and imaginary parts of the $\mathcal{PT}$-symmetric periodic potential (panel b) on that star graph. The parameters $V_0$ and $W_0$ are chosen as: $V_0=6, \; W_0=0.45$.}
\label{fig:star}
\end{figure}

We note that changing the value of the parameter $\theta$ does not affect the eigenvalues.
In the next section, we analyze the dispersion relation of branched optical lattices using the approach outlined above.

\section{Branched $\mathcal{PT}$-symmetric lattices}
\label{Branched_PT_lattices}

In this section, we extend the above study to branched $\mathcal{PT}$-symmetric lattices. The branched structure is modeled in terms of metric graphs.
We consider a simplest graph, called star graph, which has four semi-infinite edges and connected at a single vertex (see Fig.~\ref{fig:star}a).

To preserve $\mathcal{PT}$-symmetry in the system, such a graph must contain an even number of edges.
The origin of coordinates is chosen at the vertex.
For the left edges, the coordinates are fixed as $x_{-j}\in(-\infty,0]$, while for the right edges $x_j\in[0,+\infty)$, where $j=1,2$.
For simplicity, we use $x$ instead of $x_{\pm j}$, and the wavefunction on each edge is denoted by $\Psi_{\pm j}(x,t)$.

The linear Schr\"odinger equation with $\mathcal{PT}$-symmetric potential on each edge of the graph is
\begin{eqnarray}\label{LSE_star}
    i\partial_t \Psi_{\pm j}(x,t) + \partial_x^2 \Psi_{\pm j}(x,t) + V(x)\Psi_{\pm j}(x,t) = 0,
\end{eqnarray}
where the potential $V(x)$ is given in Eq.~\eqref{potential} (see Fig.~\ref{fig:star}b).
The vertex boundary conditions are imposed as wavefunction continuity and the Kirchhoff rule as
\begin{align}\label{vbc1}
\begin{split}
 &  \alpha_{-1} \Psi_{-1}(x,t)\vert_{x=0}= 
   \alpha_{-2} \Psi_{-2} (x,t)\vert_{x=0}=
   \alpha_{1} \Psi_{1}(x,t)\vert_{x=0}=
   \alpha_{2} \Psi_{2}(x,t)\vert_{x=0}, \\
 &  \frac{1}{\alpha_{-1}} \partial_x \Psi_{-1}(x,t)\vert_{x=0} + 
    \frac{1}{\alpha_{-2}} \partial_x \Psi_{-2}(x,t)\vert_{x=0}  
    = 
    \frac{1}{\alpha_1} \partial_x \Psi_1 (x, t)\vert_{x=0} +
    \frac{1}{\alpha_2} \partial_x \Psi_2(x,t)\vert_{x=0},
\end{split}
\end{align}
where $\alpha_{\pm j}$ are real constants.

Since the potential $V(x)$ satisfies the $\mathcal{PT}$-symmetry condition $V(x)=V^*(-x)$, the Hamiltonian on the star graph is $\mathcal{PT}$-symmetric.
However, the boundary conditions in Eq.~\eqref{vbc1} involve edge-dependent weights $\alpha_{\pm j}$, therefore global $\mathcal{PT}$-symmetry requires the boundary conditions to be invariant under $\mathcal{PT}$ transformation.
Consequently, the wavefunctions must satisfy the weighted $\mathcal{PT}$-symmetry relation
\begin{eqnarray}\label{pt_cond}
\alpha_{j}\Psi_{j}(x,t)
=\alpha_{-j}\Psi_{-j}^*(-x,-t),
\end{eqnarray}
which ensures global $\mathcal{PT}$ invariance on the graph.

Using Eq.~\eqref{pt_cond}, one derives the following constraint for the weights:
\begin{eqnarray}\label{sum_rule}
    \frac{1}{\alpha_{-1}^2}+
    \frac{1}{\alpha_{-2}^2}=
    \frac{1}{\alpha_{1}^2}+
    \frac{1}{\alpha_{2}^2}.
\end{eqnarray}

Similarly to the previous section, we factorize the time and coordinate as
\begin{eqnarray*}\label{Psi1}
    \Psi_{\pm j}(x,t)=e^{-i\mu t}\psi_{\pm j}(x)
\end{eqnarray*}
which yields the stationary equation
\begin{eqnarray}\label{LSE4}
    \mu \psi_{\pm j}(x)+\partial_x^2 \psi_{\pm j}(x)+V(x)\psi_{\pm j}(x)=0.
\end{eqnarray}
The plane-wave expansion on the graph is
\begin{eqnarray}\label{Psi3}
    \psi_{\pm j}(x)=\sum_{n=-N}^{N} {\varphi_n^{(\pm j)} e^{i(k+G_n)x}}.
\end{eqnarray}

\begin{figure}[t!]
    \centering
    \includegraphics[width=0.6\linewidth]{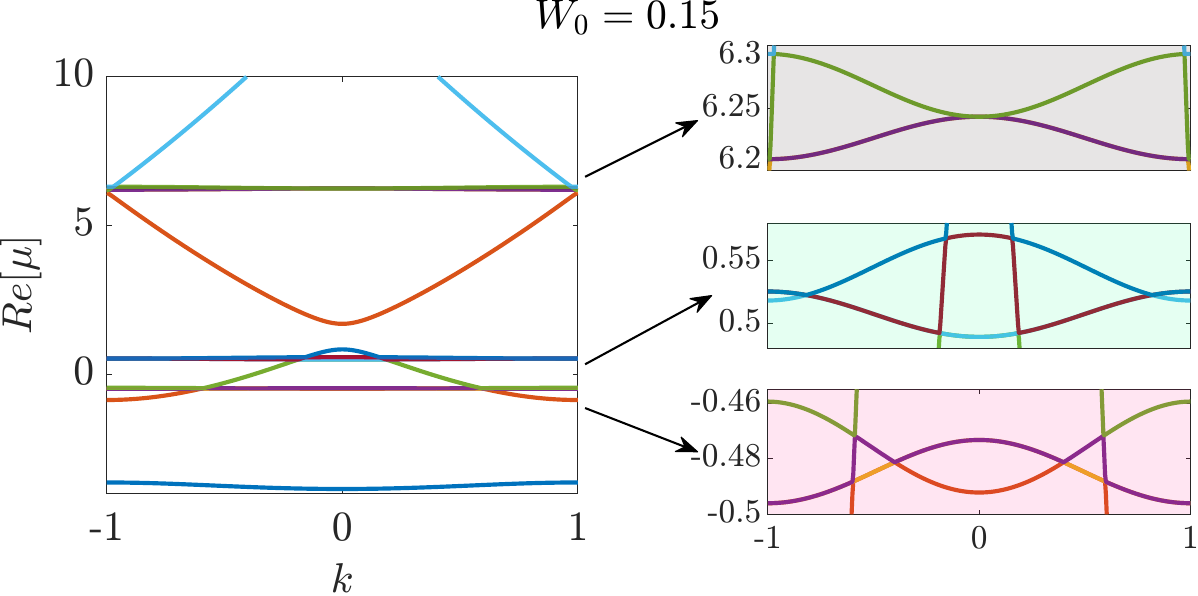}
    \caption{The dispersion relation of the $\mathcal{PT}$-symmetric star-like branched optical lattice is presented for the constraint in Eq.~\eqref{sum_rule} is satisfied, for the gain-loss parameter is below the phase transition point, i.e., the imaginary parts of the eigenvalues are zero. Other parameters are specified in the text.
}
    \label{fig:W0=0.15_fulfilled}
\end{figure}

\begin{figure}[t!]
    \centering
    \includegraphics[width=0.6\linewidth]{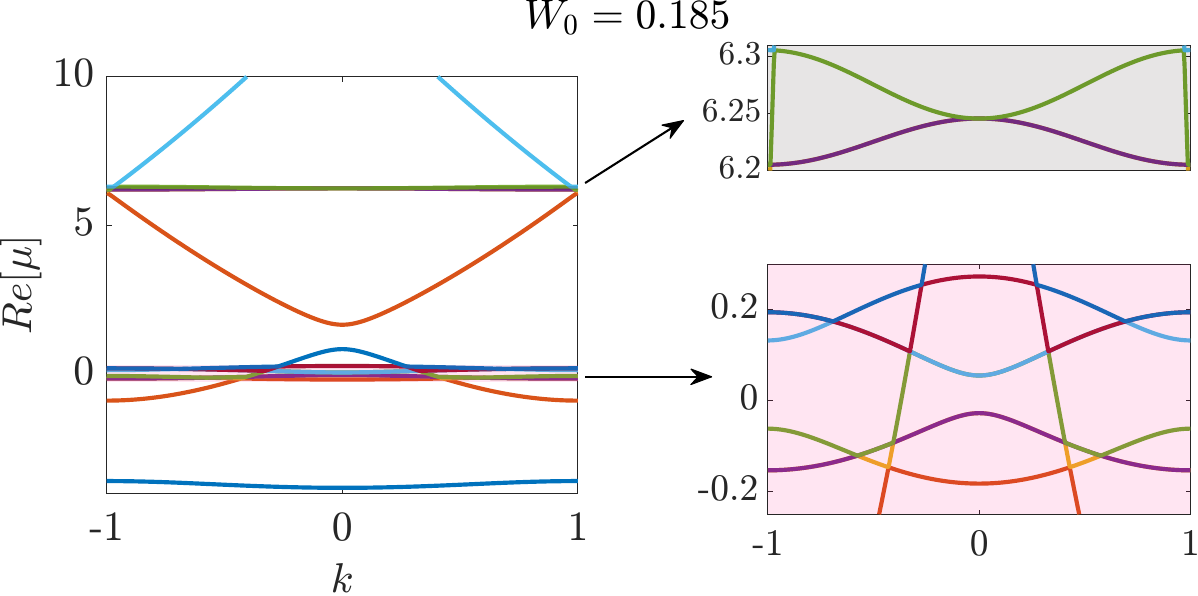}
    \caption{The dispersion relation of the $\mathcal{PT}$-symmetric star-like branched optical lattice is presented for the constraint in Eq.~\eqref{sum_rule} is satisfied. The gain-loss parameter in the vicinity of the phase transition point. The eigenvalues have zero imaginary parts. Other parameters are specified in the text.
}
    \label{fig:W0=0.185_fulfilled}
\end{figure}

\begin{figure}[t!]
    \centering
    \includegraphics[width=0.6\linewidth]{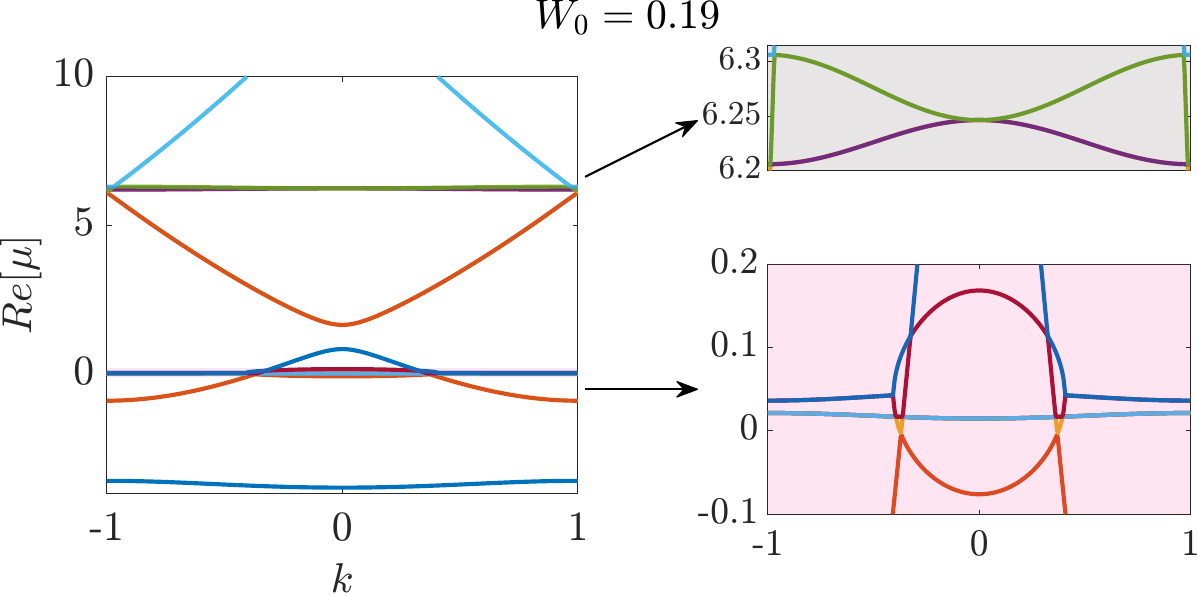}\\[1ex]
    \includegraphics[width=0.3\linewidth]{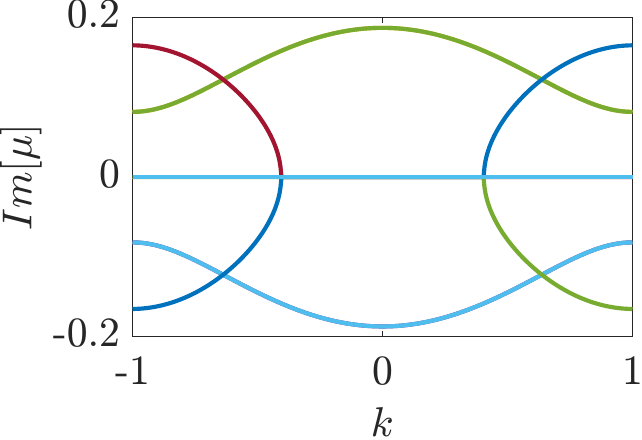}
    \caption{The dispersion relation of the $\mathcal{PT}$-symmetric star-like branched optical lattice is presented for the constraint in Eq.~\eqref{sum_rule} is satisfied. The gain-loss parameter is above the phase transition point, and in this settings, eigenvalues exhibit nonzero imaginary parts. Other parameters are specified in the text.
}
    \label{fig:W0=0.19_fulfilled}
\end{figure}

Projecting onto $\varphi_n^{(\pm j)}$ leads to the global eigenvalue problem
\begin{eqnarray}\label{eig4}
    \widetilde{H}(k) \varphi = \mu \varphi,
\end{eqnarray}
where
\begin{align*}
   & \varphi=(\varphi^{(-1)},\varphi^{(1)},\varphi^{(-2)},\varphi^{(2)})^T, \quad 
   \varphi^{(\pm j)} = (\varphi_{-N}^{(\pm j)},...,\varphi_{N}^{(\pm j)})^T,\\
   & \widetilde{H}(k) = \text{diag}(H^{(-1)}(k), H^{(1)}(k), H^{(-2)}(k), H^{(2)}(k)).
\end{align*}
The number of elements in each $\varphi^{(\pm j)}$ is $(2N+1)$, so the full vector $\varphi$ has the length $(8N+4)$.
Thus, $\widetilde{H}(k)$ is an $(8N+4)\times(8N+4)$ matrix.
Each block matrix $H^{(\pm j)}(k)$ has the same entries as $H(k)$ in Eq.~\eqref{H1}.

Using Eq.~\eqref{Psi3}, vertex boundary conditions become
\begin{align*}
\begin{split}
    \alpha_{-1} \sum_{n=-N}^{N} \varphi_n^{(-1)}=
    \alpha_{+1} \sum_{n=-N}^{N}  \varphi_n^{(+1)} &
    =
    \alpha_{-2} \sum_{n=-N}^{N}  \varphi_n^{(-2)}=
    \alpha_{+2} \sum_{n=-N}^{N} \varphi_n^{(+2)}, \\
    \frac{1}{\alpha_{-1}}\sum_{n=-N}^{N}(k+G_n) \varphi_n^{(-1)}+
    \frac{1}{\alpha_{-2}} & \sum_{n=-N}^{N} (k+G_n) \varphi_n^{(-2)} \\
    = \frac{1}{\alpha_{+1}} \sum_{n=-N}^{N} & (k+G_n) \varphi_n^{(+1)}+
    \frac{1}{\alpha_{+2}}\sum_{n=-N}^{N} (k+G_n) \varphi_n^{(+2)}.
\end{split}
\end{align*}
These equations are algebraic linear system of equations that can be rewritten as
\begin{eqnarray}\label{vbc4}
    C(k)\varphi=0.
\end{eqnarray}
To incorporate the boundary conditions, we express
\begin{eqnarray}\label{phi}
    \varphi=\Omega(k)\tilde\varphi.
\end{eqnarray}
where the columns of $\Omega(k)$ span the null space of $C(k)$.
Thus, $C(k)\Omega(k)=0$, ensuring that the boundary conditions are automatically satisfied.

\begin{figure}[t!]
    \centering
    \includegraphics[width=0.6\linewidth]{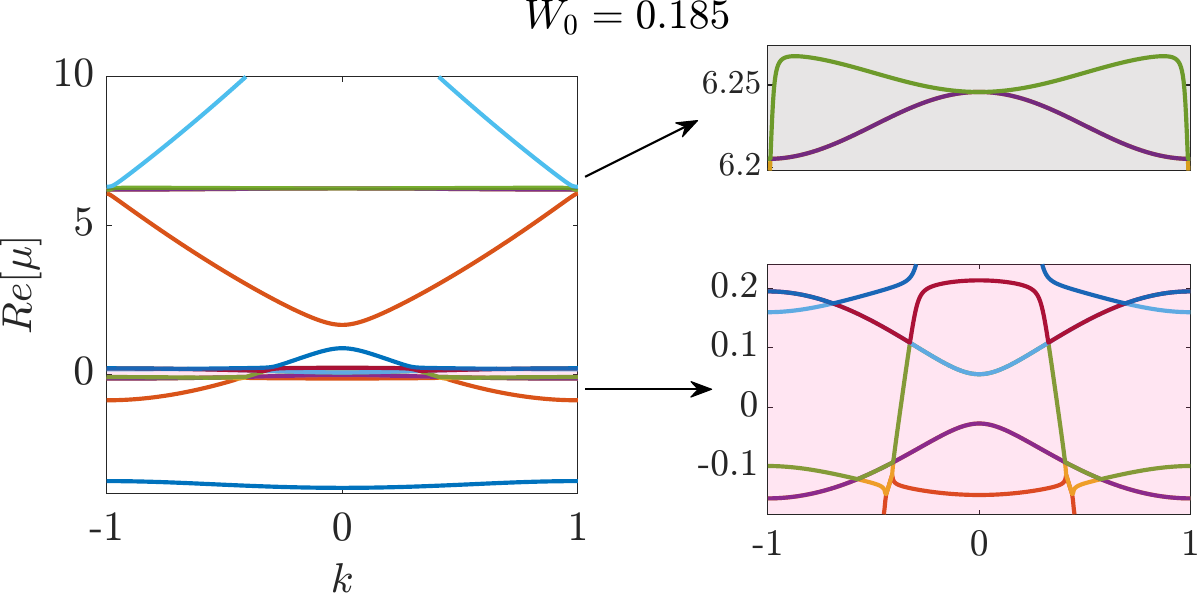}
    \\[1ex]
    \includegraphics[width=0.6\linewidth]{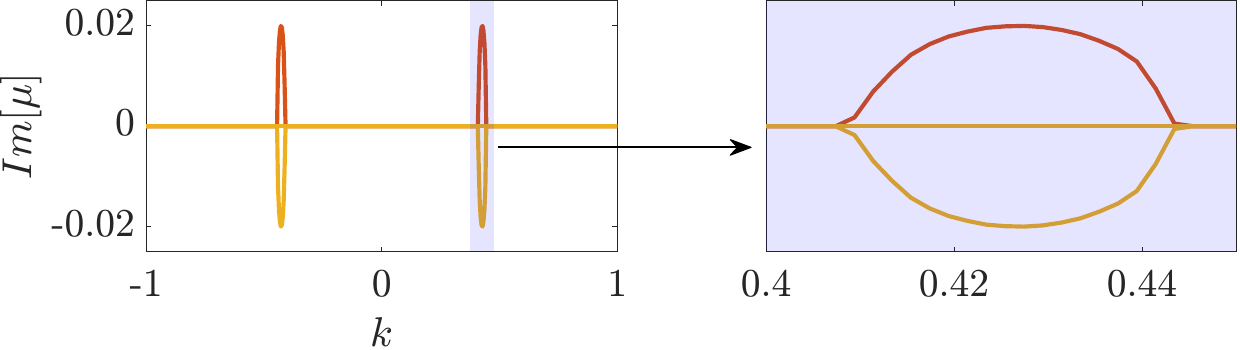}
    \caption{The dispersion relation of the $\mathcal{PT}$-symmetric star-like branched optical lattice is presented for the constraint in Eq.~\eqref{sum_rule} is broken. Other parameters are specified in the text.
}    \label{fig:W0=0.185_broken}
\end{figure}

Substituting Eq.~\eqref{phi} into Eq.~\eqref{eig4} yields
\begin{eqnarray}\label{eig5}
    \widetilde{\mathcal{H}}(k) \tilde\varphi = \mu \tilde\varphi,
\end{eqnarray}
where $\widetilde{\mathcal{H}}(k)=\Omega^{-1}(k)\widetilde H(k)\Omega(k)$.

Numerically solving this eigenvalue problem for each $k$ varying in $[-1,1]$ leads to get the dispersion relation $\mu=\mu(k)$ on the graph.

Figs.~\ref{fig:W0=0.15_fulfilled}, \ref{fig:W0=0.185_fulfilled}, \ref{fig:W0=0.19_fulfilled} and \ref{fig:W0=0.185_broken} describe the dispersion relation on the graph for the parameters given by $V_0=6, \; N=100$.
The constraint, which is determined in terms of the weight $\alpha_{\pm j}$ in Eq.~\eqref{sum_rule}, is fulfilled in Figs.~\ref{fig:W0=0.15_fulfilled}, \ref{fig:W0=0.185_fulfilled}, \ref{fig:W0=0.19_fulfilled} with $\alpha_{\pm j}=1$, and broken in Fig.~\ref{fig:W0=0.185_broken} with $\alpha_{-j} = 1$ and $\alpha_{j} = 2$, where $j=1,2$.

In all these figures, additional curves are appeared comparing with the unbranched case~\cite{Nixon} due to the branching topology of the system.
Two couple of curves are far away from each other for the gain-loss parameter $W_0<0.185$ in Fig.~\ref{fig:W0=0.15_fulfilled}.
While gain-loss parameter approaching the phase transition point $W_0\to 0.185$, the distance between two couple of curves is decreasing. Fig.~\ref{fig:W0=0.185_fulfilled} presents the dispersion relation at the phase transition point $W_0=0.185$. 
Bifurcation of the curves in the real part leads to a nonzero imaginary part of the eigenvalues for $W_0>0.185$ (see Fig.~\ref{fig:W0=0.19_fulfilled}). 
From Fig.~\ref{fig:W0=0.185_broken}, one can easily see that the eigenvalues become complex at the phase transition point $W_0=0.185$ when the sum rule is broken.

\section{Solitons in $\mathcal{PT}$-symmetric lattices}
\label{NLSE_on_line}

In this section, by following Ref.~\cite{Nixon}, we briefly recall the nonlinear
Schr\"{o}dinger equation on a line with a $\mathcal{PT}$-symmetric periodic potential in the form
\begin{eqnarray}\label{nlse4}
i\partial_t \Psi(x,t) + \partial_x^2 \Psi(x,t) + V(x)\Psi(x,t) 
+ \sigma\,\lvert \Psi(x,t)\rvert^2 \Psi(x,t) = 0
\end{eqnarray}
where $\sigma=\pm 1$ and the potential $V(x)$ is given in
Eq.~\eqref{potential}.
The solution of Eq.~\eqref{nlse4} can be analysed by employing the separation of variables as
\begin{eqnarray}\label{ansatz}
    \Psi(x,t) = e^{-i\mu t} \psi(x),
\end{eqnarray}
where $\mu$ is the propagation constant. Substituting Eq.~\eqref{ansatz} into
Eq.~\eqref{nlse4} yields
\begin{eqnarray}\label{nlse2}
    \mu \psi(x)+\partial^2_x \psi(x) + V(x) \psi(x) + \sigma\vert \psi(x) \vert^2 \psi(x) = 0.
\end{eqnarray}
Eq.~\eqref{nlse2} together with the periodic potential \eqref{potential} cannot
be solved analytically. Ref.~\cite{Nixon} proposed numerical approaches based on the
squared-operator iteration method and the Newton-conjugate-gradient method.

To study the stability of a solution $\psi(x)$, we introduce a small
perturbation of the form~\cite{Nixon}
\begin{equation}\label{perturb}
    \Psi(x,t)=e^{-i\mu t} \left[ \psi(x)+u(x)e^{\lambda t}+v^*(x)e^{\lambda^*t} \right],
\end{equation}
where $\lvert u(x)\rvert , \lvert v(x)\rvert \ll \lvert \psi(x)\rvert$. Substitution of
Eq.~\eqref{perturb} into Eq.~\eqref{nlse4} and linearization gives the eigenvalue
problem
\begin{equation}
i\mathcal{L} \begin{pmatrix} u \\ v \end{pmatrix} = \lambda \begin{pmatrix} u \\ v \end{pmatrix},
\end{equation}
where
\begin{align*}
& \mathcal{L} = \begin{pmatrix}
L_{1} & L_{2} \\
-L_{2}^* & -L_{1}^*
\end{pmatrix}, \quad
L_{1} = \mu + \partial_{x}^2 + V(x) + 2\sigma |\psi|^2, \;\;
L_{2} = \sigma \psi^2.
\end{align*}
A soliton is linearly unstable if $\lambda$ 
has nonzero positive real part. Numerical
calculations of both stable and unstable, fundamental and dipole $\mathcal{PT}$-symmetric solitons were
reported in Ref.~\cite{Nixon}.

\begin{figure}[t!]
    \centering
    \begin{minipage}{0.4\textwidth}
        \centering
        \includegraphics[width=0.99\linewidth]{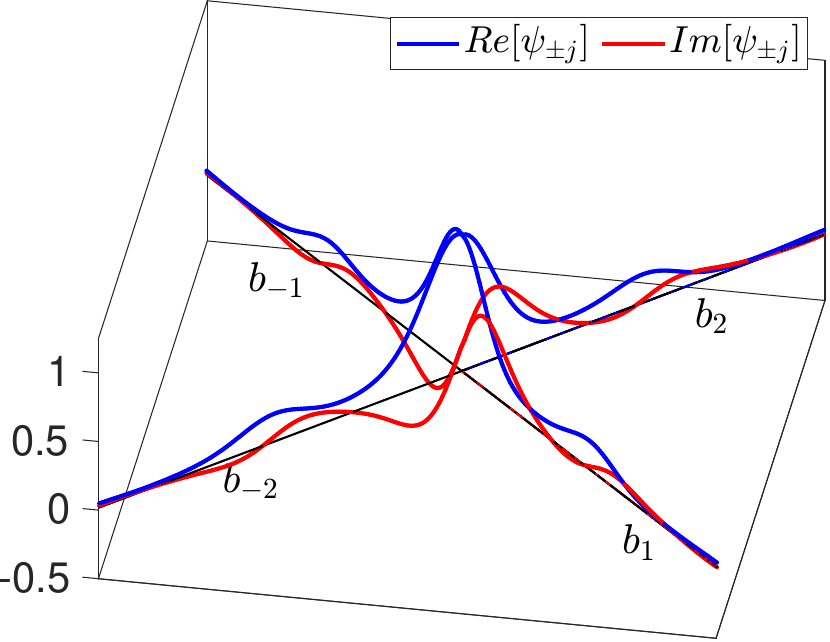}
        \small{(a)}
    \end{minipage}
    \begin{minipage}{0.4\textwidth}
        \centering
        \includegraphics[width=0.99\linewidth]{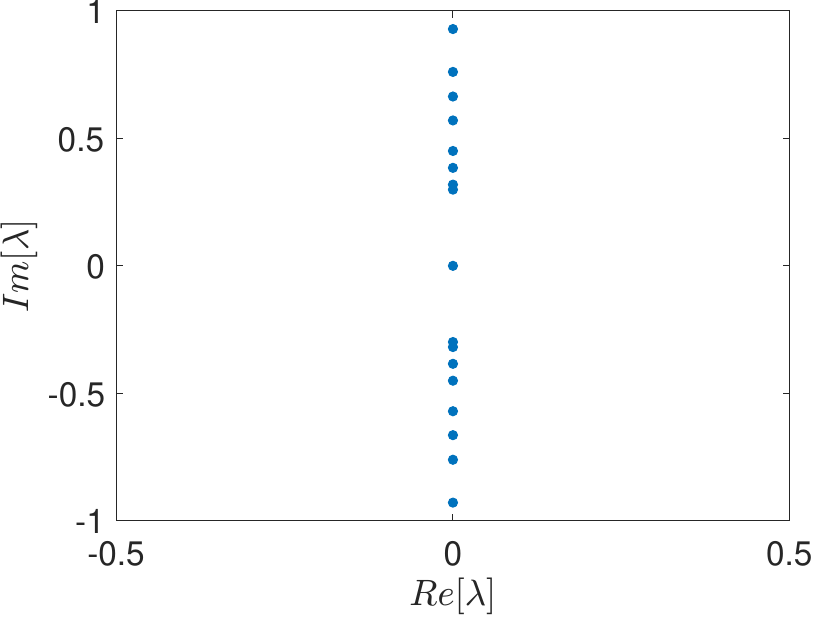}
        \small{(b)}
    \end{minipage}
\caption{$\mathcal{PT}$-symmetric soliton profiles (panel a) and the corresponding stability spectra (panel b) on the graph for the sum rule in Eq.~\eqref{sum_rule1} is fulfilled with $\beta_{\pm j}=1$.
Other parameters are chosen as
$V_0 = 6,\; W_0 = 0.45,\; \mu = -3.5$.}
\label{fig:stable1}
\end{figure}

\section{Solitons in $\mathcal{PT}$-symmetric branched lattices}
\label{NLSE_on_star}

We now extend the above analysis to a star graph with four semi-infinite edges (see
Fig.~\ref{fig:star}a). The nonlinear Schr\"{o}dinger equation on the graph is written as
\begin{equation}\label{nlse1_star}
    i\partial_t \Psi_{\pm j}(x,t) + \partial^2_x \Psi_{\pm j}(x,t) + V(x) \Psi_{\pm j}(x,t)
    + \beta_{\pm j}\vert \Psi_{\pm j}(x,t) \vert^2 \Psi_{\pm j}(x,t) = 0,
\end{equation}
where $j=1,2$, $\beta_{\pm j}$ denotes the edge-dependent nonlinearity coefficient, and
$V(x)$ is the same $\mathcal{PT}$-symmetric periodic potential as given in Eq.~\eqref{potential}.
Here, we restrict our analysis to the focusing regime, i.e., the nonlinearity coefficients $\beta_{\pm j}$ are assumed to be positive on all edges.

\begin{figure}[t!]
    \centering
    \begin{minipage}{0.4\textwidth}
        \centering
        \includegraphics[width=0.99\linewidth]{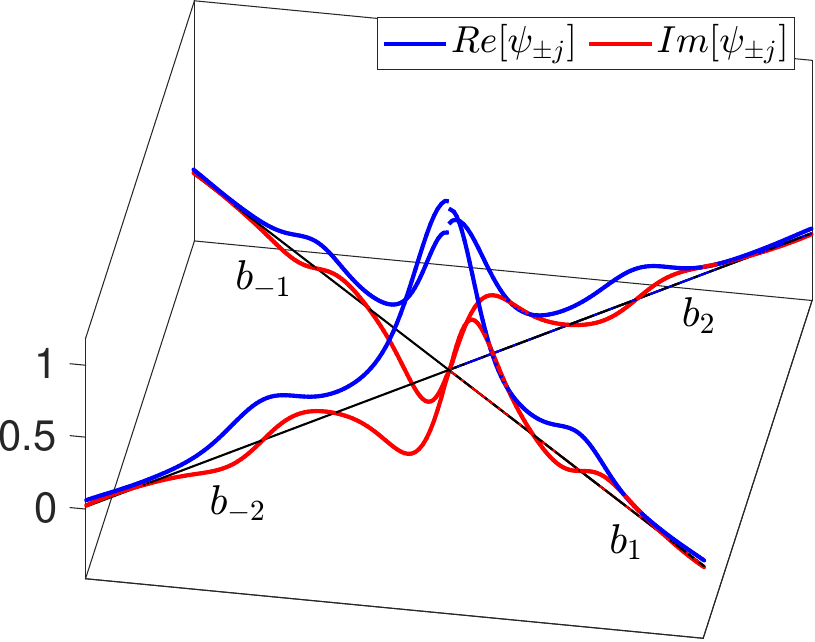}
        \small{(a)}
    \end{minipage}
    \begin{minipage}{0.4\textwidth}
        \centering
        \includegraphics[width=0.99\linewidth]{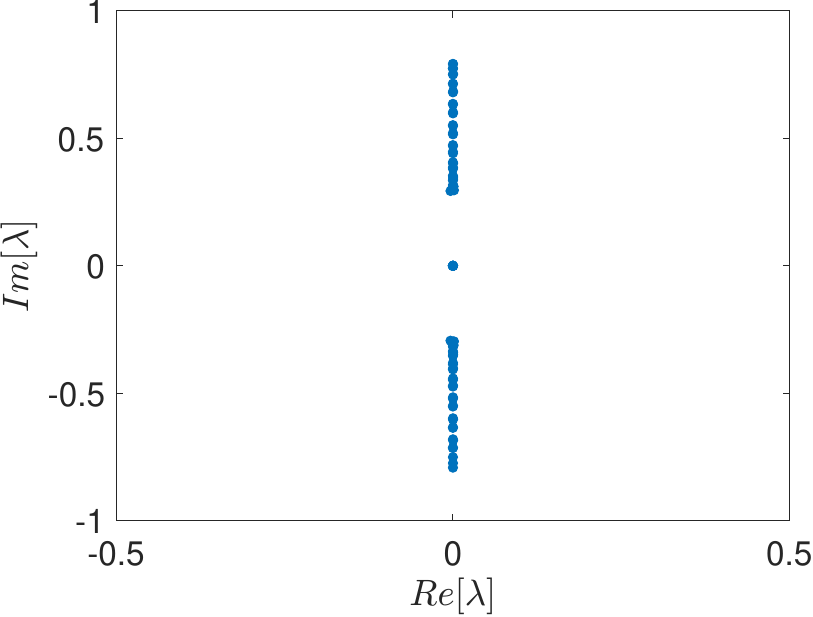}
        \small{(b)}
    \end{minipage}
    \caption{$\mathcal{PT}$-symmetric soliton profiles (panel a) and the corresponding stability spectra (panel b) on the graph for the sum rule in Eq.~\eqref{sum_rule1} is fulfilled with $\beta_{-1}=10/12, \; \beta_{-2}=10/8, \; \beta_{1}=10/11, \; \beta_{2}=10/9$.
Other parameters are the same as the Fig.~\ref{fig:stable1}.}
\label{fig:stable2}
\end{figure}

The vertex boundary conditions are the same as in
Eq.~\eqref{vbc1}. In the nonlinear regime the weight parameters $\alpha_{\pm j}$ can be determined in terms of $\beta_{\pm j}$.

The $\mathcal{PT}$-symmetry on the graph requires
\begin{eqnarray}\label{pt_cond1}
\alpha_{j} \Psi_{j}(x,t)=
\alpha_{-j} \Psi_{-j}^*(-x,-t).
\end{eqnarray}

Substituting Eq.~\eqref{pt_cond1} into the vertex boundary
conditions~\eqref{vbc1} yields to obtain the constraint as
\begin{eqnarray} \label{sum_rule1}
    \frac{\alpha_{\pm j}}{\alpha_1} = \sqrt{\frac{\beta_{\pm j}}{\beta_1}}, \quad
    \frac{1}{\beta_{-1}}+\frac{1}{\beta_{-2}} = \frac{1}{\beta_1} + \frac{1}{\beta_2}.
\end{eqnarray}

Using the ansatz
\begin{eqnarray}
    \Psi_{\pm j}(x,t) = e^{-i\mu t} \psi_{\pm j}(x)
\end{eqnarray}
Eq.~\eqref{nlse1_star} reduces to
\begin{eqnarray}\label{nlse2_star}
    \mu \psi_{\pm j}(x)+\partial^2_x \psi_{\pm j}(x) + V(x) \psi_{\pm j}(x)
    + \beta_{\pm j} \vert \psi_{\pm j}(x) \vert^2 \psi_{\pm j}(x) = 0.
\end{eqnarray}
The boundary conditions at the vertex become
\begin{align*}
\begin{split}
   \alpha_{-1} \psi_{-1}(x)\vert_{x=0}=
   & \alpha_{-2} \psi_{-2}(x)\vert_{x=0}
   =\alpha_{1}  \psi_{1}(x)\vert_{x=0}=
   \alpha_{2} \psi_{2}(x)\vert_{x=0}, \\
   \frac{1}{\alpha_{-1}} \partial_x \psi_{-1}(x)\vert_{x=0} & + 
    \frac{1}{\alpha_{-2}} \partial_x \psi_{-2}(x)\vert_{x=0} 
    =\frac{1}{\alpha_1} \partial_x \psi_1(x)\vert_{x=0}+
    \frac{1}{\alpha_2} \partial_x \psi_2(x)\vert_{x=0},
\end{split}
\end{align*}
Eq.~\eqref{nlse2_star} with the vertex boundary conditions above has no closed-form analytical
solution. To solve it numerically, we apply the FFT (Fast Fourier transform) to the Eq.~\eqref{nlse2_star}. By introducing
\[
    \widetilde{\psi}_{\pm j}(x)
    = -\mathcal{F}^{-1}\!\left[k^2\, \mathcal{F}(\psi_{\pm j}(x))\right],
\]
Eq.~\eqref{nlse2_star} becomes
\begin{eqnarray}\label{nlse20_star}
    \mu \psi_{\pm j}(x)+ \widetilde{\psi}_{\pm j}(x) + V(x) \psi_{\pm j}(x) 
    + \beta_{\pm j} \vert \psi_{\pm j}(x) \vert^2 \psi_{\pm j}(x) = 0,
\end{eqnarray}

We split the Eq.~\eqref{nlse20_star} into a system of coupled nonlinear algebraic equations for real and imaginary parts as
\begin{align*}
\begin{split}
    \mu \psi_{\pm j}^{(r)}(x)+\widetilde{\psi}^{(r)}_{\pm j}&(x) + [ \text{Re}[V(x)] \psi_{\pm j}^{(r)}(x)-\text{Im}[V(x)] \psi_{\pm j}^{(i)}(x) ]  + \beta_{\pm j} [ (\psi_{\pm j}^{(r)}(x))^2 + (\psi_{\pm j}^{(i)}(x))^2  ] \psi_{\pm j}^{(r)}(x) = 0, \\
    \mu \psi_{\pm j}^{(i)}(x)+\widetilde{\psi}^{(i)}_{\pm j}&(x) + [ \text{Re}[V(x)] \psi_{\pm j}^{(i)}(x)+\text{Im}[V(x)] \psi_{\pm j}^{(r)}(x) ] 
     + \beta_{\pm j} [ (\psi_{\pm j}^{(r)}(x))^2 + (\psi_{\pm j}^{(i)}(x))^2  ] \psi_{\pm j}^{(i)}(x) = 0,
\end{split}
\end{align*}
where $\psi_{\pm j}^{(r)}(x)$, $\widetilde{\psi}^{(r)}_{\pm j}(x)$, and $\psi_{\pm j}^{(i)}(x)$, $\widetilde{\psi}^{(i)}_{\pm j}(x)$ are real and imaginary parts of $\psi_{\pm j}(x)$ and $\widetilde{\psi}_{\pm j}(x)$, respectively.
Similarly, vertex boundary conditions take the following form for $\psi_{\pm j}^{(r)}(x)$ and $\psi_{\pm j}^{(i)}(x)$
\begin{align*}
\begin{split}
   \alpha_{-1} \psi_{-1}^{(r)}(x)\vert_{x=0}=
   & \alpha_{-2} \psi_{-2}^{(r)}(x)\vert_{x=0}
   = \alpha_1 \psi_1^{(r)}(x)\vert_{x=0}=
   \alpha_2 \psi_2^{(r)}(x)\vert_{x=0}, \\
   \frac{1}{\alpha_{-1}} \partial_x \psi_{-1}^{(r)}(x)\vert_{x=0} & + 
    \frac{1}{\alpha_{-2}} \partial_x \psi_{-2}^{(r)}(x)\vert_{x=0} 
    = \frac{1}{\alpha_1} \partial_x \psi_1^{(r)}(x)\vert_{x=0}+
    \frac{1}{\alpha_2} \partial_x \psi_2^{(r)}(x)\vert_{x=0}, \\
   \alpha_{-1} \psi_{-1}^{(i)}(x)\vert_{x=0}=
   &\alpha_{-2} \psi_{-2}^{(i)}(x)\vert_{x=0}
   = \alpha_1 \psi_1^{(i)}(x)\vert_{x=0}=
   \alpha_2 \psi_2^{(i)}(x)\vert_{x=0}, \\
   \frac{1}{\alpha_{-1}} \partial_x \psi_{-1}^{(i)}(x)\vert_{x=0} & + 
    \frac{1}{\alpha_{-2}} \partial_x \psi_{-2}^{(i)}(x)\vert_{x=0}
     = \frac{1}{\alpha_1} \partial_x \psi_1^{(i)}(x)\vert_{x=0}+
    \frac{1}{\alpha_2} \partial_x \psi_2^{(i)}(x)\vert_{x=0}.
\end{split}
\end{align*}

\begin{figure}[t!]
    \centering
    \begin{minipage}{0.4\textwidth}
        \centering
        \includegraphics[width=0.99\linewidth]{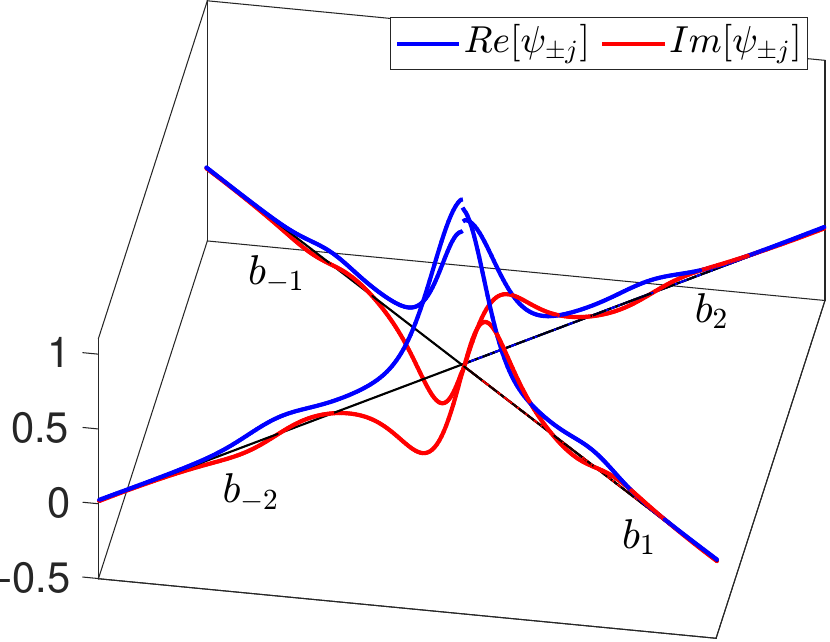}
        \small{(a)}
    \end{minipage}
    \begin{minipage}{0.4\textwidth}
        \centering
        \includegraphics[width=0.99\linewidth]{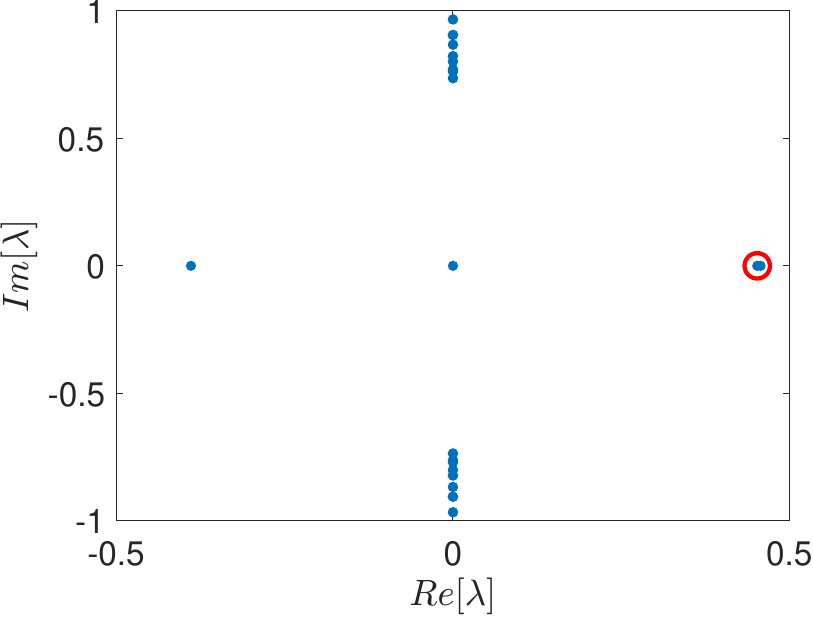}
    \small{(b)}
    \end{minipage}
    \caption{$\mathcal{PT}$-symmetric soliton profiles (panel a) and the corresponding stability spectra (panel b) on the graph for the sum rule in Eq.~\eqref{sum_rule1} is fulfilled with $\beta_{-1}=10/12, \; \beta_{-2}=10/8, \; \beta_{1}=10/11, \; \beta_{2}=10/9$.
Other parameters are chosen as: $V_0=6, \; W_0=0.55, \; \mu=-3.5$.}
\label{fig:unstable3}
\end{figure}

\begin{figure}[t!]
    \centering
    \begin{minipage}{0.5\textwidth}
        \centering
        \includegraphics[width=0.99\linewidth]{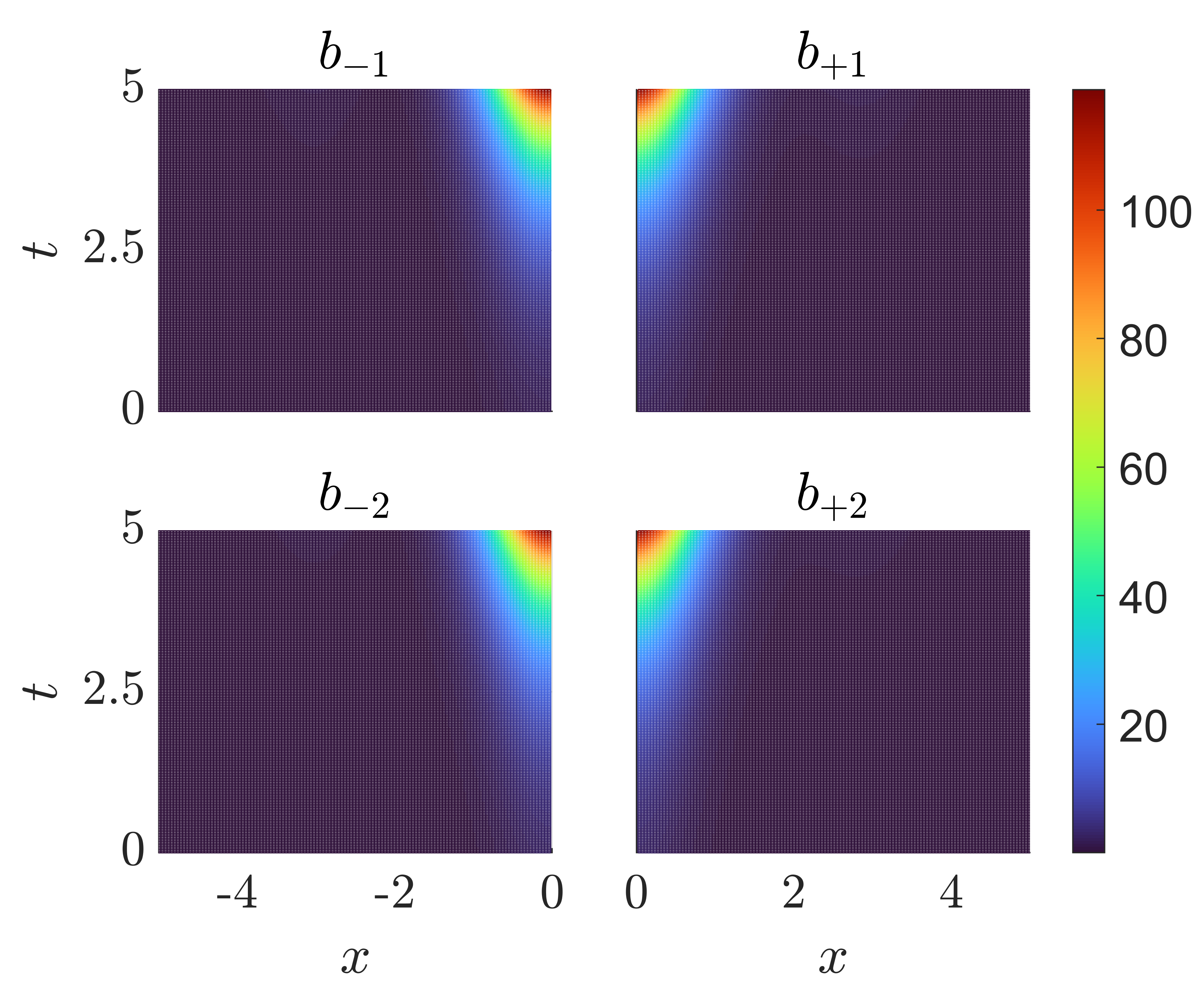}
        \small{(a)}
    \end{minipage}
    \begin{minipage}{0.4\textwidth}
        \centering
        \includegraphics[width=0.99\linewidth]{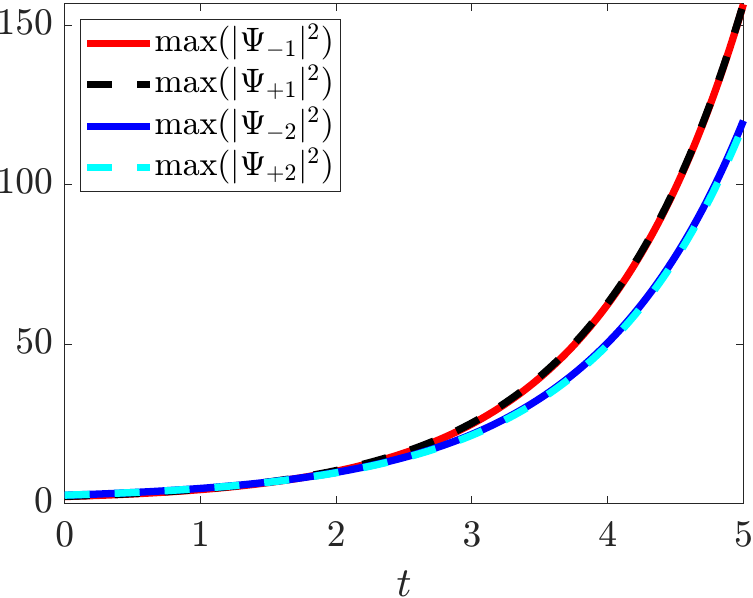}
        \small{(b)}
    \end{minipage}
    \caption{Evolution of field intensity $|\Psi_{\pm j}|^2$ (panel a) and time dependence its maximum value (panel b) of unstable PT solitons on the each edge of the graph.
    }
\label{fig:exp}
\end{figure}

The above nonlinear algebraic system of equations with the boundary conditions at the vertex can be solved by means of the package’s built-in iterative solver, which is based on a Newtonian iteration scheme.
For the initial guess, we have chosen the hyperbolic secant function whose center of the profile is located at the vertex.

The same manner is employed to analyse the stability of $\mathcal{PT}$-symmetric solitons on the graph by perturbing the solution, as demonstrated in the Refs.~\cite{Nixon,  Musslimani, Bakirtas1, Old} as
\begin{eqnarray}\label{perturb2}
    \Psi_{\pm j}(x,t) = e^{-i\mu t} \left[ \psi_{\pm j}(x) + e^{\lambda t}u_{\pm j}(x)+e^{\lambda^* t}v^*_{\pm j}(x) \right],
\end{eqnarray}
where $\vert u_{\pm j}(x) \vert,\vert v_{\pm j}(x) \vert \ll \vert \psi_{\pm j}(x) \vert$.
Substituting the Eq.~\eqref{perturb2} into Eq.~\eqref{nlse1_star}, and linearizing them with respect to higher order powers of $u_{\pm j}(x)$ and $v_{\pm j}(x)$ gives the following eigenvalue problem
\begin{eqnarray}
    i\mathcal{L} \omega = \lambda \omega,
\end{eqnarray}
where $\omega$ and the operator $\mathcal{L}$ are defined as
\begin{align*}
\begin{split}
    & \omega(x) = \text{diag}\left( \omega_{-1}(x),\omega_{-2}(x),\omega_1(x), \omega_2(x) \right), \quad
    \omega_{\pm j}(x)=(u_{\pm j}(x),v_{\pm j}(x))^T, \\[1ex]
    & \mathcal{L} = \text{diag}\left( \mathcal{L}_{-1},\mathcal{L}_{-2},\mathcal{L}_1, \mathcal{L}_2 \right), 
    \quad
    \mathcal{L}_{\pm j} = \begin{pmatrix}
        \mathcal{L}^{(1)}_{\pm j} & \mathcal{L}^{(2)}_{\pm j} \\[1ex]
        -\mathcal{L}^{(2)*}_{\pm j} & -\mathcal{L}^{(1)*}_{\pm j}
    \end{pmatrix}
    \\[1ex]
    & \mathcal{L}^{(1)}_{\pm j} = \mu +\partial_x^2 +V(x) +2\beta_{\pm j}\vert \psi_{\pm j}(x) \vert^2,\quad
    \mathcal{L}^{(2)}_{\pm j} = \beta_{\pm j} \psi_{\pm j}^2(x).
\end{split}
\end{align*}
The same vertex boundary conditions in Eq.~\eqref{vbc1} are applied for $\omega_j(x)$ as
\begin{align*} 
\begin{split}
    \alpha_{-1} \omega_{-1}(x)\vert_{x=0}=
   &\alpha_{-2}  \omega_{-2} (x)\vert_{x=0}
   =\alpha_{1} \omega_{1}(x)\vert_{x=0}=
   \alpha_{2} \omega_{2}(x)\vert_{x=0}, \\
   \frac{1}{\alpha_{-1}} \partial_x \omega_{-1}(x)\vert_{x=0} & + 
    \frac{1}{\alpha_{-2}} \partial_x \omega_{-2}(x)\vert_{x=0}    
    =\frac{1}{\alpha_1} \partial_x \omega_1(x)\vert_{x=0}+
    \frac{1}{\alpha_2} \partial_x \omega_2(x)\vert_{x=0}.
\end{split}
\end{align*}

Analogously, the Fourier transform method is employed for the eigenvalue problem with the boundary conditions at the vertex to obtain the stability spectra.

We numerically solve the nonlinear Schr\"odinger equation with the boundary conditions at the vertex and demonstrate the soliton profiles on each edge of the graphs for different parameters
in Figs.~\ref{fig:stable1}, \ref{fig:stable2} and \ref{fig:unstable3} (panel a). Corresponding stability spectra are shown on panel b on these figures.

For the stable solitons, identical values of the nonlinearity coefficient $\beta_{\pm j}$ lead to identical amplitudes in Fig.~\ref{fig:stable1}, and
non-equal values of $\beta_{\pm j}$ cause different amplitudes of stable solitons in Fig.~\ref{fig:stable2}, which can be clearly seen at the vertex.
Fig.~\ref{fig:unstable3} shows the switching of stable solitons to unstable ones while crossing the gain-loss parameter at the phase transition point $W_0=0.5$.

The eigenvalue highlighted with red circle in Fig.~\ref{fig:unstable3}b determines the associated unstable $\mathcal{PT}$-symmetric soliton. The corresponding spatial–temporal dynamics of this soliton on each edge of the graph together with the time evolution of maximum value of field intensity $|\Psi_{\pm j}(x,t)|^2$ in \eqref{perturb2} are shown in Fig.~\ref{fig:exp}.
From this plot one can easily observe that the soliton is blowing up for the gain loss parameter at $W_0=0.55$.
Unlike in the linear regime, the phase transition point shifts from 0.185 to 0.5, due to the cubic nonlinearity.
In our numerical experiment, we have examined the stability spectra for the gain-loss parameter around $W_0=0.5$ and chose the limited set of parameters.
The power as a function of the chemical potential reveals that the cubic nonlinearity induces a significant shift in the $\mathcal{PT}$-symmetry breaking threshold.


Violating the sum rule in Eq.~\eqref{sum_rule1} causes the breaking of $\mathcal{PT}$-symmetry in the system.
Interestingly, in our numerical experiment, we note that our system is highly sensitive to the values of nonlinearity coefficient $\beta_{\pm j}$. Even a small deviation from this constraint leads to the disappearance of the soliton profile.
The disappearance of the soliton profile can be attributed to the two main factors:
first, breaking the $\mathcal{PT}$-symmetry in the system and second, the pronounced sensitivity of the solution to variations in the nonlinearity coefficients.
Therefore, in all calculations, we have chosen the values of $\beta_{\pm j}$ to be close to unity.

\section{Conclusions}\label{conclusion}

In this work, we have studied $\mathcal{PT}$-symmetric branched optical lattices by employing the nonlinear Schrödinger equation with a $\mathcal{PT}$-symmetric periodic potential on metric graphs.
In the unbranched case, i.e., on a line, we have proved mathematically the condition for the gain–loss parameter using the plane-wave expansion method, showing that 
the similarity transformed matrix Hamiltonian is Hermitian under the condition~\eqref{cond}.
By modelling the branched structure in terms of metric graphs,
the dispersion relation is obtained for different parameters on the star graph with four edges.
We have shown that the exceptional point (or phase-transition point) can be controlled using the constraint in Eq.~\eqref{sum_rule}, i.e., when the system is below the $\mathcal{PT}$-symmetry-breaking threshold, this constraint provides real eigenvalues by choosing appropriate values of the weight coefficients $\alpha_{\pm j}$, and all eigenvalues become real.

In the nonlinear regime, we have numerically solved the cubic nonlinear Schrödinger equation with the same $\mathcal{PT}$-symmetric periodic potential on the star graph.
The linear stability of solitons is analyzed by adding perturbation terms to the complex amplitude of the field.
When the constraint in Eq.~\eqref{sum_rule1} is broken, we have noted that the amplitude of the soliton becomes nearly zero.

The achievement of this work comes from the fact that we have obtained tunable $\mathcal{PT}$-symmetric optical lattices, where the tuning relies on the constraints in Eqs.~\eqref{sum_rule} and \eqref{sum_rule1}.
This tunability allows one to regulate the spectral properties, phase-transition threshold, and stability of the solitons. Such a feature makes the proposed model a useful theoretical platform for designing and analysing $\mathcal{PT}$-symmetric nonlinear wave systems in complex optical networks.
Moreover, this property of such systems lays the foundation for the application of $\mathcal{PT}$-symmetry in real optical devices, such as $\mathcal{PT}$-symmetric optical couplers, unidirectional optical isolators, and $\mathcal{PT}$-symmetric optical switches, among others.

\end{document}